\documentclass[english,prd,twocolumn,superscriptaddress,nofootinbib,preprintnumbers]{revtex4}
\usepackage[latin1]{inputenc}
\usepackage{graphicx}
\usepackage{amssymb}
\usepackage{color}

\usepackage{amsmath}
\usepackage{amsfonts}
\usepackage{dcolumn}
\usepackage{hyperref}

\def\be{\begin{equation}}
\def\ee{\end{equation}}

\makeatother

\usepackage{babel}
\makeatother
\begin{document}

\title{FLRW cosmology in Weyl-Integrable Space-Time}

\author{Radouane Gannouji}
\affiliation{Department of Physics, Faculty of Science, Tokyo University of Science,
1-3, Kagurazaka, Shinjuku-ku, Tokyo 162-8601, Japan}

\author{Hemwati Nandan}
\affiliation{Department of Physics, Gurukula Kangri Vishwavidayalaya, Haridwar 249404, India}

\author{Naresh Dadhich}
\affiliation{IUCAA, Post Bag 4, Ganeshkhind, Pune 411 007, India}

\begin{abstract}

We investigate the Weyl space-time extension of general relativity
(GR) for studying the FLRW cosmology through focusing and defocusing
of the geodesic congruences. We have derived the equations of
evolution for  expansion, shear and rotation in the Weyl space-time.
In particular, we consider the Starobinsky modification,
$f(R)=R+\beta R^2-2\Lambda$, of gravity in the Einstein-Palatini
formalism, which turns out to reduce to the Weyl integrable
space-time (WIST) with the Weyl vector being a gradient. The
modified Raychaudhuri equation takes the form of the Hill-type
equation which is then analysed to study the formation of the
caustics. In this model, it is possible to have a Big Bang
singularity free cyclic Universe but unfortunately the periodicity
turns out to be extremely short.
\end{abstract}

\maketitle

\section{Introduction}

In order to solve the problems (viz. singularity, flatness, horizon)
concerning the early universe in cosmology, the modification of the classical
geometrical structure of General Relativity (GR) becomes pertinent and various
such modifications are attempted time and again \cite{Hawking:1973uf,Wald:1984rg, Joshi:1987wg,
  Ellis:1971pg, Carroll:1997ar}. Various inflationary cosmological models
\cite{Tasi}, scalar-tensor theories and the models resulting
from the string theory \cite{Fujii:2003pa} such as dilaton gravity are considered to
solve the above-mentioned  problems which are still not satisfactorily answered.
These modifications amount in general to introducing a scalar field for some models while
for others they appeal to a unified description of the fundamental forces for understanding cosmology
of the early universe \cite{Albrecht:1982wi}. In the context of unification of the forces, the Weyl
space-time through its non-metric connection with a vector field $(W_\mu)$ could facilitate unification of
the two long range forces, electromagnetic and gravity. \\

The field $W_\mu$ defines an enlarged structure of space-time namely
the Weylian manifold. The Weyl extension is a natural generalized
space-time structure if we employ light rays and freely falling
observers instead of clocks \cite{clock}, \cite{Ehlers} as the basic
tools for maping the space-time. 
the EPS axiomatization shown \cite{Ehlers} that Weyl space-time is the natural structure if we use light
rays and freely falling particles instead of clocks as basic tools.
In this approach, the Riemannian geometry is replaced by a Weyl geometry as the structure of the
space-time which enlarges the framework of the GR space-time and as a direct
consequence, a purely geometric fundamental new field is introduced into the theory
of gravitation.
The characterisation of the space-time and geodesic
congruences with
Weyl geometry in the Einstein-Palatini
formalism need to be considered carefully for the complete manifestation of the gravitational fields \cite{Poulis:2011ve,Fatibenea:2011aw}.\\
The formation of caustics \cite{Poisson,Anirvan} in a given space-time is important to
examine the singularities of the space-time foliations \cite{Joshi:1987wg,Sardanashvily}.
The caustics is an undesirable feature as it indicates failure of proper evolution of physical fields in certain region of space-time \cite{Felder,Goswami}. A caustic is therefore a singularity of a geodesic congruence where
the equation of motion becomes untenable at such points \cite{Poisson}. The study of the formation
of caustics in the field configuration relative a given space-time background is therefore very important and the analysis of the Raychaudhuri
equation \cite{Poisson, Sayan1, Dadhich} in diverse various settings \cite{Anirvan, Anirvan1} plays a crucial role in probing the physical viability of a model in relation to the occurrence of the space-time singularities \cite{Joshi:1987wg, Goswami}.  \\

In the present paper, we study the formation of caustics by examining focusing and defocusing of geodesic congruences in the Weyl space-time which becomes WIST for $f(R)$ gravity in the Einstein-Palatini formalism. It is then equivalent to a conformal transformation. One of the basic tools of analysis is however the Raychaudhuri equation. The paper is organized as follows: In the next section, we set up the Weyl space-time structure which is followed by the computation of the kinematic parameters namely expansion scalar, shear and rotation with $(1+3)$ space-time split in section III. In Sec IV, we set up the Einstein-Palatini formalism for $f(R)$ gravity followed by the study of formation of caustics in various cases by using the Raychaudhuri equation which is recasted as a Hill-type equation in section V. Finally, we end up with the conclusion.

\section{Weyl Space-Time}
In the Weyl geometry, the connection is related to the metric by the Weyl vector ($W_\mu$) and is defined by
\be
\label{Weylspace}
\nabla_\rho g_{\mu\nu}=W_{\rho}g_{\mu\nu}
\ee

where $ \nabla $ is a covariant derivative in the Weyl manifold.
This equation is invariant under the conformal
transformations  (i.e.  gauge transformation) \cite{Poberii:1994rz}

\begin{equation}
\begin{split}
\label{gauge}
g_{\mu \nu} &\rightarrow  e^{ w} g_{\mu \nu}\\
W_{\mu} &\rightarrow W_{\mu} +\partial_{\mu} w.
\end{split}
\end{equation}

Using equation (\ref{Weylspace}), one can write the connection ($\Gamma^\rho_{\mu\nu}$) in the following form

\be
\Gamma^\rho_{\mu\nu}=\bar{\Gamma}^\rho_{\mu\nu}+\Delta^\rho_{\mu\nu}
\ee

where $\bar{\Gamma}^\mu_{\nu\rho}$ is the Levi-Civita connection and $\Delta^\mu_{\nu\rho}$ is an additional term which depends completely on the Weyl vector, the so-called deviation tensor.

For the torsion free connection, it also follows from the equation (\ref{Weylspace}) that

\be
\Delta^\rho_{\mu\nu}=-\frac{1}{2}\left(\delta^\rho_\mu W_\nu+\delta^\rho_\nu W_\mu-g_{\mu\nu} W^\rho\right).
\ee

Because of this additional vector, the Riemann tensor constructed from this connection is not antisymmetric in the last two indices. Therefore the Einstein-Maxwell theory can be described by the second Ricci tensor $ R^{~~~~\rho}_{\mu \nu \rho } $, also known as the homothetic curvature tensor or the segmental curvature tensor ($ V $) \cite{Schouten} as given below,
\begin{align}
V_{\mu \nu}&=\frac{1}{4}g^{\rho \sigma}R_{ \mu \nu \rho \sigma}=\frac{1}{2} \left( \partial_{\mu} W_{\nu} - \partial_{\nu} W_{\mu}\right)\nonumber\\
&=\frac{1}{2} F_{\mu\nu}.
\end{align}

Here $ F_{\mu\nu} $ is the electromagnetic field tensor for the Weyl vector $ W_{\mu} $  which serves as the electromagnetic four potential.

The Lagrangian in the classical electromagnetism can therefore be written as a
geometrical term $(\frac{1}{4}F_{\mu\nu}F^{\mu \nu} \equiv V_{\mu
  \nu}V^{\mu \nu})$. But as pointed out by Einstein, such description
will also lead to a second clock effect (or a twin paradox of the second kind)
which may cause inconsistencies \cite{Brown:1999fx,Novello:1992tb,arXiv:0912.0432} unless the tensor $ F_{\mu \nu} $ vanishes identically; i.e.

\be
\label{WIST}
\partial_{\mu} W_{\nu}-\partial_{\nu} W_{\mu}=0.
\ee

It can not describe the electromagnetic field anymore, but it has
various interesting cosmological consequences which will be of our concern in the rest of the paper. The equation (\ref{WIST}) has the obvious solution for $W_\mu$ as a  gradient of a scalar, $W_\mu=-\partial_\mu \phi$) \cite{Weyl,Novello:2008ra}. This space-time is known as Weyl Integrable Space-Time (WIST) or reducible Weylian space-time.

We  will consider this geometry for present work because it is the only viable
subclass of Weyl Space-Time \cite{Pauli}. The structure is the same as that of the
conformal transformation of the metric $g_{\mu\nu}$, the metric which defines
the Levi-Civita connexion. In fact, if we define the metric ($f$) in the form

\be
f_{\mu\nu}=e^{\phi} g_{\mu\nu},
\ee

the connection $\Gamma^\rho_{\mu\nu}$ is now the Levi-Civita connection of the conformal metric $f_{\mu\nu}$. This is why the Weyl-Integrable space-times are also referred as conformally-Riemannian \cite{Salim:1996ei}.

It is then easy to calculate the Riemann and the Ricci tensors corresponding to the conformal metric and are given as below,

\begin{widetext}
\begin{align}
\label{eq:Riemann}
R_{\mu\nu\rho}^{~~~~\sigma}(f)&=\bar{R}_{\mu\nu\rho}^{~~~~\sigma}(g)+\delta^\sigma_{[\mu}\bar{\nabla}_{\nu]}\bar{\nabla}_\rho\phi-g_{\rho[\mu}\bar{\nabla}_{\nu]}\bar{\nabla}^\sigma \phi
+\frac{1}{2}\delta^\sigma_{[\nu} \bar{\nabla}_{\mu]}\phi \bar{\nabla}_\rho \phi
-\frac{1}{2}g_{\rho[\nu} \bar{\nabla}_{\mu]}\phi \bar{\nabla}^\sigma \phi
-\frac{1}{2}g_{\rho[\mu}\delta^\sigma_{\nu]}\bar{\nabla}_\alpha \phi \bar{\nabla}^\alpha\phi\\
\label{eq:Ricci}
R_{\mu\nu}(f)&=\bar{R}_{\mu\nu}(g)-\bar{\nabla}_{\mu\nu}\phi-\frac{1}{2}g_{\mu\nu}\bar{\Box}\phi+\frac{1}{2}\bar{\nabla}_\mu \phi \bar{\nabla}_\nu\phi-\frac{1}{2}g_{\mu\nu}\bar{\nabla}_\rho\phi\bar{\nabla}^\rho\phi
\end{align}
\end{widetext}

where $\bar{\nabla}_\mu$ is the covariant derivative for the Levi-Civita connection $\bar{\Gamma}$ of the metric $g_{\mu\nu}$.\\

\section{$(1+3)$-decomposition and the kinematic quantities}

The study of the evolution of geodesic congruences essentially hinges on the evolution of the deviation vector  $\xi^\mu$ connecting the two neighbouring geodesics in the congruence. Let $u^\mu$ be a tangent vector to the congruence of timelike geodesics and so we write ${\cal L}_u \xi= {\cal L}_\xi u =0$ which means that $u^\nu\nabla_\nu\xi^{\mu} = \xi^\nu \nabla_\nu u^{\mu}$. This leads to
\begin{align}
\frac{\rm d}{\rm dt} u^\mu\xi_\mu &\equiv u^\beta\nabla_\beta (u^\mu\xi_\mu) \nonumber \\
&= \xi_\mu u^\beta u^\sigma\left(\Delta_{\beta\sigma}^\mu-\Delta_{\beta\sigma}^\mu\right) \nonumber
\\
&= 0
\end{align}
and thus $\xi^\mu$ is orthogonal to $u^\mu$ everywhere. \\

Since the congruence is associated with the timelike vector field $u^\mu$, one can decompose the space-time metric into longitudinal and transverse parts with the projection tensor $h$ associated with the unit vector $u$ in the following form,

\be
\label{}
h_{\mu\nu}\equiv g_{\mu\nu}+u_\mu u_\nu.
\ee

We now have the fully orthogonally projected covariant derivative $D$ defined by

\be
D_\mu T^{\alpha_1 \cdots}_{\qquad ~~\beta_1 \cdots}=h_\mu^{~\nu} h^{\alpha_1}_{~\rho_1}\ldots h_{\beta_1}^{~\sigma_1} \ldots \nabla_\nu T^{\rho_1 \cdots}_{\qquad~~ \sigma_1 \cdots}.
\ee

It is easy to show that $D_\mu h_{\alpha \beta}=h_{\mu}^{~\rho} W_\rho h_{\alpha \beta}$, which is a Weyl space-time defined by the projection vector of $W_\rho$ in the $3$-dimensional subspace. We can now project the tensor field $\nabla u$ in the components  orthogonal and parallel to $u$ as follows:
\be
\label{eq:decomposition}
\nabla_\mu u_\nu=D_\mu u_\nu-u_\mu \dot{u}_\nu +\frac{1}{2}\left(u_\nu W_\mu+u_\mu u_\nu \dot{W}\right)
\ee
where $\dot{W}=u^\gamma W_\gamma=-u^\gamma \nabla_\gamma \phi=-\dot{\phi}$. The usual decomposition is recovered for $W_\mu=0$. It is easy to see that for a timelike geodesic ($u^\gamma \bar{\nabla}_\gamma u_\nu=0$ where $\bar{\nabla}$ is the covariant derivative relative to the Levi-Civita connection), we have

\be
\dot{u}_\nu=u^\gamma \nabla_\gamma u_\nu=-\frac{1}{2} W_\nu
\ee

taking equation (\ref{eq:decomposition}) to the form

\be
\nabla_\mu u_\nu=D_\mu u_\nu+\frac{1}{2}\left(u_\mu W_\nu+u_\nu W_\mu+u_\mu u_\nu \dot{W}\right).
\ee

It is worth noting that the Weyl vector measures the acceleration ($\dot{u}_\nu$) in the Weylian space-time. These terms would be absent only when the Weyl field is zero.

In order to have the kinematics of the geodesic congruence, one introduces the
evolution tensor $B_{\mu\nu} \equiv D_\nu u_\mu$  which measures the failure of  $\xi^\mu$ being parallely transported along the congruence. The evolution tensor on the subspace orthogonal to $u$ can be decomposed in its irreducible components (i.e. into trace, symmetric-tracefree and antisymmetric parts) as follows:
 \be
D_\nu u_\mu= \frac{1}{3}\theta h_{\mu \nu}+\sigma_{\mu\nu}+\omega_{\mu\nu} \label{decom1}
\ee
where the trace $\theta$, the symmetric part $\sigma_{\mu\nu}$ and the antisymmetric part
$\omega_{\mu\nu}$ are the expansion scalar, shear tensor (or vorticity) and rotation tensor respectively. This is the same as for the Levi-Civita connection,
\be
\bar{\nabla}_\nu u_\mu= \bar{D}_\nu u_\mu = \frac{1}{3}\bar{\theta} h_{\mu \nu}+\bar{\sigma}_{\mu\nu}+\bar{\omega}_{\mu\nu} \label{dcom2}.
\ee
After some straightforward calculations, it follows easily
\be
D_\mu u_\nu= \bar{D}_\mu u_\nu-\frac{1}{2}h_{\mu \nu} \dot{W}.
\ee
In view of the above decompositions, the three kinematic quantities in the two cases are thus related by
\begin{align}
\label{def:theta}
\theta&=\bar{\theta}-\frac{3}{2}\dot{W}\\
\label{def:rotation}
\sigma_{\mu\nu}&=\bar{\sigma}_{\mu\nu}\\
\label{def:shear}
\omega_{\mu\nu}&=\bar{\omega}_{\mu\nu}
\end{align}
Note that it is only the expansion scalar which is sensitive to the Weyl scalar field while the other two shear and vorticity remain unaffected. However the evolution equations would be affected because the equations for shear and rotation also involve the expansion. \\
The equation (\ref{def:theta}) reduces to the trivial solution in the case of a WIST. In fact, in a FLRW spacetime with the redefinition of the scale factor as $a^2\rightarrow e^{\phi}a^2$, we have $\theta\rightarrow \theta+3\dot\phi/2$.

Now in order to derive the Raychaudhuri equation, we begin with the evolution equation for $B_{\mu\nu}$ itself,

\begin{widetext}
\be
\label{eq:evolution}
u^\gamma \nabla_\gamma B_{\mu\nu}=
-B_{\mu\gamma}g^{\gamma\sigma}B_{\sigma\nu}
+u^\gamma u_\sigma R_{\gamma\nu\mu}^{~~~~\sigma}
+u_{\mu}u^{\gamma}V_{\nu\gamma}
-\frac{1}{2}h_\mu^{~\alpha}h_{\nu}^{~\beta}\left(\nabla_{\beta}W_{\alpha}-\frac{1}{2}W_{\alpha}W_{\beta}\right)
-\frac{W^{\alpha}}{2}\left( B_{\alpha \nu}u_{\mu}+ B_{\mu\alpha} u_{\nu}\right)
\ee
\end{widetext}

where

\be
h_{\nu}^{~\alpha}h_{\mu}^{~\beta}u^{\gamma}R_{\gamma\alpha\beta}^{~~~~\sigma}u_{\sigma}=u^{\gamma}u_{\sigma}R_{\gamma\nu\mu}^{~~~~\sigma}+u_{\mu}u^{\gamma}V_{\nu\gamma}.
\ee
It is worth noting that the equation (\ref{eq:evolution}) takes the usual GR form $u^\gamma \nabla_\gamma B_{\mu\nu}=-B_{\mu\gamma}g^{\gamma\sigma}B_{\sigma\nu}+u^\gamma u_\sigma R_{\gamma\nu\mu}^{~~~~\sigma}$ in the absence of Weyl vector $W_\mu$. Since $u^\gamma B_{\mu\nu}\nabla_\gamma g^{\mu\nu}=-\theta \dot{W}$ and using the equation (\ref{eq:evolution}), the evolution equations for the three kinematic parameters are given by
\begin{widetext}
\begin{align}
\label{eq:expansion}
u^\gamma \nabla_\gamma \theta&=
-\frac{1}{3}{\theta}^2
-{\sigma}^2
+{\omega}^2
-R_{\mu\nu}u^\mu u^\nu
-\frac{1}{2}h^{\mu\nu}\left(\nabla_{\mu}W_\nu-\frac{1}{2}W_{\mu}W_{\nu}\right)
-\theta\dot{W}\\
\label{eq:rotation}
u^\gamma \nabla_\gamma \sigma_{\mu\nu}&=
-\frac{2}{3}\theta\sigma_{\mu\nu}
-\sigma_\mu^{~\alpha}\sigma_{\alpha\nu}
-\omega_\mu^{~\alpha}\omega_{\alpha\nu}
+\frac{1}{3}h_{\mu\nu}\left(\sigma^2-\omega^2\right)
+\frac{1}{3}h_{\mu\nu}R_{\alpha\beta}u^\alpha u^\beta
+u^\gamma u_\sigma R_{\gamma\nu\mu}^{~~~~\sigma}\nonumber\\
&-\frac{1}{2}\left(\nabla_{(\alpha}W_{\beta)}-\frac{1}{2}W_\alpha W_\beta\right) \left(h_\mu^{~\alpha}h_\nu^{~\beta}-\frac{1}{3}h_{\mu\nu}h^{\alpha\beta}\right)
-W^{\alpha}\sigma_{\alpha(\mu}u_{\nu)}
+\frac{1}{2}\left(V_{\mu\nu}+2u^{\gamma}u_{(\mu}V_{\nu)\gamma}\right)\\
\label{eq:shear}
u^\gamma \nabla_\gamma \omega_{\mu\nu}&=
-\frac{2}{3}\theta\omega_{\mu\nu}
+2 \sigma_{\gamma[\mu}w_{\nu]}^{~\gamma}
+W^{\alpha}\omega_{\alpha[\mu}u_{\nu]}
\end{align}
\end{widetext}
where $R_{\mu\nu(\rho\sigma)}=V_{\mu\nu}g_{\rho\sigma}$. Note that the usual part i.e. as in GR and the Weyl affected part separate out neatly in the above equations. Recall that the evolution equation for expansion, and similarly for the shear and rotation, the Raychaudhuri equation for the Levi-Civita connection reads as
\be
\label{eq:RayGR}
u^\gamma \bar{\nabla}_\gamma \bar{\theta}=
-\frac{1}{3}{\bar{\theta}}^2
-{\bar{\sigma}}^2
+{\bar{\omega}}^2
-\bar{R}_{\mu\nu}u^\mu u^\nu
\ee
For the dynamical evolution, the "true" variable for the expansion is $\theta$ as defined in (\ref{def:theta}) and not $\bar\theta$ and also the time variation is defined as $\dot{S}_{\mu\nu}=u^\gamma\nabla_{\gamma}S_{\mu\nu}\neq u^\gamma\bar\nabla_{\gamma}S_{\mu\nu}$, where $S$ is a tensor field. Therefore the evolution of a model defined in a Riemannian manifold with the presence of a scalar field is different from its evolution when the scalar field is included in the definition of the connection.

In the case of the WIST when the Weyl vector is a gradient, the equations (\ref{eq:expansion},\ref{eq:rotation},\ref{eq:shear}) reduce to

\begin{widetext}
\begin{align}
u^\gamma \nabla_\gamma \theta&=-\frac{1}{3}{\theta}^2-{\sigma}^2+{\omega}^2-R_{\mu\nu}u^\mu u^\nu+\frac{1}{2}h^{\mu\nu}\left(\nabla_{\mu\nu}\phi+\frac{1}{2}\nabla_{\mu}\phi\nabla_{\nu}\phi\right)+\theta\dot{\phi}\\
u^\gamma \nabla_\gamma \sigma_{\mu\nu}&=-\frac{2}{3}\theta\sigma_{\mu\nu}-\sigma_\mu^{~\alpha}\sigma_{\alpha\nu}-\omega_\mu^{~\alpha}\omega_{\alpha\nu}+\frac{1}{3}h_{\mu\nu}\left(\sigma^2-\omega^2\right)+\frac{1}{3}h_{\mu\nu}R_{\alpha\beta}u^\alpha u^\beta+u^\gamma u_\sigma R_{\gamma\mu\nu}^{~~~~\sigma}\nonumber\\
&+\frac{1}{2}\left(\nabla_{\alpha \beta}\phi
+\frac{1}{2}\nabla_\alpha\phi\nabla_\beta\phi \right) \left(h_\mu^{~\alpha}h_\nu^{~\beta}-\frac{1}{3}h_{\mu\nu}h^{\alpha\beta}\right)+
\sigma_{\alpha(\mu}u_{\nu)}\nabla^\alpha\phi\\
u^\gamma \nabla_\gamma \omega_{\mu\nu}&=-\frac{2}{3}\theta\omega_{\mu\nu}+2 \sigma_{\gamma[\mu}w_{\nu]}^{~\gamma}-\omega_{\alpha[\mu}u_{\nu]}\nabla^\alpha\phi.
\end{align}
\end{widetext}
In the particular case of a Friedmann Universe (i.e. in absence of shear and rotation), the Raychaudhuri equation (\ref{eq:expansion}) for the expansion scalar assumes the simple form,

\be
\label{eq:expansionscalar}
\dot\theta+\frac{\theta^2}{3}-\frac{\theta}{2}\dot\phi=-R_{\mu\nu}u^\mu u^\nu
\ee
which indeed reduces to the second  Friedmann equation with $\theta\equiv \bar\theta = 3 \dot a/a$ (where $a$ is scale factor) in the standard cosmology in absence of the scalar field. We emphasize that by considering the equations (\ref{eq:Ricci},\ref{def:theta},\ref{eq:RayGR}), we recover the equation (\ref{eq:expansionscalar}) and the same would be true for the shear and rotation as well.

\section{Einstein-Palatini Formalism}

In the preceding section, we have derived the equation for the geodesic congruences in the case of a Weyl Integrable Space-Time.  So far we have not specified the Weyl field $ \phi $ and in fact it could be left as a free function which defines a new structure of the space-time or alternatively it could be derived on variation of the action in the Einstein-Palatini formalism (EPF). In that case it is equivalent to fixing the gauge (\ref{gauge}).\\

In fact all the actions of the following form in the EPF
\be
\label{eq:epfaction}
S=\int {\rm d^4 x}\sqrt{-g} f(R,\psi,X),
\ee
where $\psi$ is a scalar field and
\be
X=-\frac{1}{2}g^{\mu \nu}\partial_\mu \psi \partial_\nu \psi
\ee
are equivalent to a gravitational theory in a WIST. In the EPF, we do not fix any metricity condition as given by equation (\ref{WIST}) and therefore it is not necessary to add a constraint in the action as it is done in the variational principle with constraints \cite{Cotsakis:1997cj}. The equations of motion are given by
\begin{align}
f_{,R}R_{(\mu\nu)}-\frac{1}{2}g_{\mu \nu} f-\frac{1}{2}\partial_\mu \psi \partial_\nu \psi f_{,X}&=T_{\mu \nu}\,,\\
f_{,\psi}+2W_{\mu}\psi^{,\mu}f_{,X}+\nabla_{\mu}\Bigl(f_{,X}\psi^{,\mu}\Bigr)&=0\,,\\
\nabla_\alpha (\sqrt{-g}f_{,R} g^{\mu \nu})&=0.
\end{align}
It then readily determines the Weyl scalar as
\be
\phi=\ln f_{,R}
\ee
which seems to show that the variational principle has selected a WIST. However, this is not a WIST but a Riemann space-time with an undetermined gauge in the vacuum. What it says is that when matter is introduced, the structure of the space-time can no longer remain Riemannian but it becomes Weylian. This is an interesting feature that emerges from the consideration of the  Einstein-Palatini formalism.

\section{Evolution of the Raychaudhuri equation in $f(R)$-gravity}

Let us consider matter as a perfect fluid described by the energy-momentum tensor,
\be
T_{\mu \nu}=(\rho+P)u_\mu u_\nu+P g_{\mu\nu}.
\ee
We assume that the matter fields are minimally coupled to the gravitational field ($g$) and the conservation equation with the Levi-Civita connection $\bar{\Gamma}$ is
\be
\bar{\nabla}^\mu T_{\mu \nu}=0,
\ee
which leads to
\be
\label{eq:barconservation}
\dot{\rho}+\bar{\theta} (\rho+P)=0.
\ee
Using equation (\ref{def:theta}), it takes the form

\be
\label{eq:conservation}
\dot{\rho} +\theta (\rho+P)=\frac{3}{2} (\rho+P)\dot{\phi}.
\ee

For $f(R)$ gravity, the equation (\ref{eq:conservation}) writes

\begin{align}
\label{eq:conservation2}
\dot\rho +\theta\rho &\frac{2MF(1+w)}{2MF-\rho(1+w)(1-3w)}\nonumber\\
&+\frac{3(1+w)\dot w}{2MF-\rho(1+w)(1-3w)}\rho^2
=0
\end{align}

where $w=P/\rho$, $F\equiv f'(R)$ and $M\equiv\frac{1}{3}\Bigl(\frac{f'(R)}{f''(R)}-R\Bigr)$. $R$ is solution of the equation

\be
\label{eq:tracefr}
f'(R)R-2f(R)=-\rho(1-3w).
\ee

Also the Raychaudhuri equation reduces to

\begin{align}
\label{eq:expansionscalar2}
\dot\theta&+\frac{\theta^2}{3}\frac{2MF}{2MF-\rho(1+w)(1-3w)}\nonumber\\
&+\theta\rho\frac{\dot w}{2MF-\rho(1+w)(1-3w)}=-\frac{\rho-f/2}{F}
\end{align}

We will consider a particular model where $f(R)=R+\beta R^2-2\Lambda$ which then readily determines the Weyl scalar as
\be
\phi=\ln (\alpha+2\beta \rho),\quad \text{with}~~\alpha=1+8\beta\Lambda
\ee

where we assumed the presence of dust ($w=0$).

The conservation and Raychaudhuri equation (\ref{eq:conservation2},\ref{eq:expansionscalar2}) read respectively as

\begin{align}
\label{eq:rhoexp}
\dot{\rho}+\theta \rho \frac{\alpha+2 \beta \rho }{\alpha-\beta  \rho}&=0\,,\\
\label{eq:exp}
\dot{\theta}+\frac{\theta^2}{3}\frac{\alpha+2\beta\rho}{\alpha-\beta\rho}&=-\frac{(1-8\beta\Lambda)\rho-\beta\rho^2-2\alpha\Lambda}{2(\alpha+2\beta\rho)}.
\end{align}

The standard results follow for $\beta=0$.

We will consider only the physical case where $(\beta,\Lambda)$ are positive which implies the positivity of $\alpha$. Therefore by redefining the variable $\theta=3\gamma(\rho)\frac{\dot G}{G}$, where
$\gamma$ is a positive function of $\rho$, one can rewrite the equation (\ref{eq:exp}) in the following Hill-type equation,

\be
\ddot G+\frac{(2-\alpha)\rho-\beta\rho^2-2\Lambda\alpha}{6(\alpha+2\beta\rho)\gamma(\beta \rho)}G=0
\ee

The caustic  formation would occur due to focusing when $G=0$ and $\dot G<0$ at a finite time \cite{Anirvan}. It is clear from the above equation that it is the sign of $(2-\alpha)\rho-\beta\rho^2-2\Lambda\alpha$ which will play the critical role. The special case where the cosmological constant is null simplifies the problem. In that case, for $\beta\rho<1$ it would be the usual GR result of focusing to the formation of caustics ultimately leading to the singularity. On the other hand, if however $\beta\rho>1$, the caustics formation will depend on the initial conditions. The focusing is then only possible for initially $\theta_0<\theta_0^c<0$ where $\theta_0^c$ is some critical value, else there would be defocusing. While in GR it is always focussing in the absence of rotation.\\

In the case where we have a positive cosmological constant, two domains come out. One is for $\beta\Lambda>1/24$ for which we can have defocusing (depending on the initial conditions) for any density $\rho$. And the second domain is for $\beta\Lambda<1/24$ where defocusing takes place for $\beta\rho<\frac{1-8\beta\Lambda-\sqrt{1-24\beta\Lambda}}{2}$ or for $\beta\rho>\frac{1-8\beta\Lambda+\sqrt{1-24\beta\Lambda}}{2}$. We recover the previous result when $\Lambda=0$.\\

\begin{figure}
\includegraphics[scale=.7]{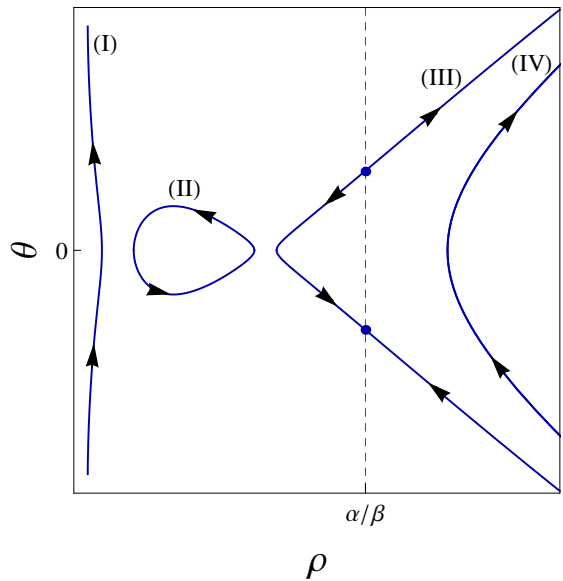}
\caption{Identification of different branches for a closed Universe in $\rho$-$\theta$ parameter plane corresponding to different generic features of the solutions of equation (\ref{eq:FLRW1}) with and without a cosmological constant.}
\label{fig:zone}
\end{figure}

\begin{figure}
\includegraphics[scale=.7]{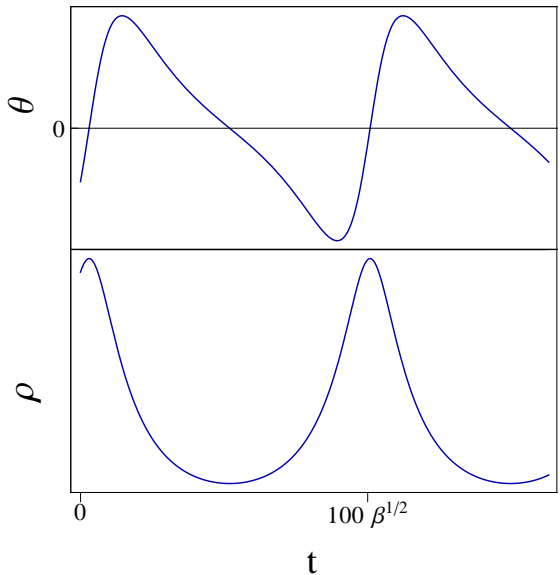}
\caption{Evolution of $\theta$ and $\rho$ for the cyclic Universe corresponding to a periodic solution.}
\label{fig:thetarhoc}
\end{figure}

In the domains where we found focusing there is always a singularity in the geodesic congruence. However in the regime where there is a possible defocusing, one can avoid the singularity. The issue of singularity is thus sensitive to the choice of initial conditions.\\
The equations (\ref{eq:rhoexp}) and (\ref{eq:exp}) after integration leads to
\be
\label{eq:FLRW1}
\theta^2=3 \Lambda+3\rho\frac{1+2\beta\Lambda+\beta\rho/2}{1+8\beta\Lambda+2\beta\rho}+C\rho^{2/3}
\ee

where C is a constant of integration which coincides with the spatial curvature (k) of the FLRW, $C=-9k/(\rho_{0}^{2/3}a_0^2)$ and the density is defined as $\rho=\rho_{0}a_{0}^3/a^{3}$. Therefore depending on the initial conditions, or equivalently on the spatial curvature, the model can lead or not to defocusing in the domains previously derived. We emphasize that there is also an other initial condition which has to be chosen as $\rho(t=0)$ or $\theta(t=0)$.\\
It is interesting to note that $\bar\theta=\theta \frac{\alpha+2\beta\rho}{\alpha-\beta\rho}$ is singular at $\beta\rho=\alpha$ while $\theta$ is not.  $\theta$ and $\bar \theta$ are equal for $\beta=0$. It is important to remember that the function  which governs the expansion is $\theta$ and not $\bar{\theta}$ and the two are however related through the equation (\ref{def:theta}). In the EPF, both metric and connection, which will however include the Weyl field, are independent variables.\\
The possible solutions of equation (\ref{eq:FLRW1}) under different conditions manifest themselves in terms of four different branches (see Fig.\ref{fig:zone}) for the evolution of expansion scalar with density. In the special case where $\Lambda=0$, we have $\theta=0$ (rebound) only for $C<0$ which means that the bounce can only be for a closed Universe $(k=1)$.
This solution corresponds to the branches (III) and (IV) as mentioned in the Fig.\ref{fig:zone}. Depending on the initial condition, the system can have two different behaviors. Let us consider, the case of collapse where initially $\beta\rho<\alpha$ and as it proceeds $\rho$ increases and $\theta$ decreases. After a finite time, the density of the system reaches the value $\rho=\alpha/ \beta$ for which the curvature is finite. This point corresponds to an attractor. In the case of GR where $\beta\rightarrow 0$, $\beta\rho=\alpha$ corresponds to $\rho=\infty$. This Universe can not have a primordial bounce except if there is a discontinuous transition to an other branch. Therefore the singularity in GR is replaced by a discontinuity in this model. For the other branch (IV) where we integrate from $\beta \rho>\alpha$, the system is continuous but as the Universe collapses the density decreases, after the bounce we will have an expansion of the Universe with an increasing density. The solution corresponding to the branch (III) can not lead to defocusing. However the solution corresponding to the branch (IV) leads to defocusing but in a Universe where the density increases during expansion (see Fig.\ref{fig:thetarhoc}) and it can not describe our Universe. In the case when $\Lambda=0$ only these two branches are present and the corresponding solutions are not satisfactory form the view point of our Universe. Therefore for the consistency of the model, a cosmological constant need to be included in it which leads to two more solutions of equation (\ref{eq:FLRW1}) presented as branch (I) and branch (II) in Fig.\ref{fig:zone}.\\
The branch (I) is a very well known result that exist even in the case where $\beta=0$. It describes a closed Universe with a cosmological constant. 

However the branch (II) represents a cyclic Universe.
Let us consider a collapse of a Universe where the initial density is considered in the branch (II). As it proceeds the density increases (see Fig.\ref{fig:thetarhoc}) and reaches the maximum. This would then lead to defocusing and the collapse will end into a rebound ($\theta=0$) rather than a caustics. This is an effect due to the $R^2$ modification together with the cosmological constant. A necessary condition for the existence of the cyclic Universe is $0<\beta\Lambda<1/24$, which implies the extreme importance of the two constants  (i.e. $\beta$ and $\Lambda$) in the existence of the cyclic Universe. After the rebound, the density decreases in an expanding Universe. At small densities, the model will have an other rebound which creates the cyclic Universe. This solution exist only for a closed Universe and has a maximum periodicity $T\propto 100\sqrt{\beta}$.  However the variation of the density of matter for the same Universe would be $\Delta\rho\propto 0.5 M_{Pl}^{2}/\beta$.
 By considering $\beta \propto H_{0}^{-2}$, the Universe will have a periodicity larger that the Hubble time but at the same tine, the density of matter will have an extremely small variation $\Delta\rho/\rho_0\propto 1$ where $\rho_0$ is the density today.

\section{Conclusion}
One of the natural generalizations of the Riemannian nature of space-time is the Weyl space-time. Ehlers {\it et al.} \cite{Ehlers} had already shown that the Weyl space-time is indeed the most general structure for the study of gravitational field as a geometric field. Such generalizations are quite pertinent in addressing the present day observational challenges posed in terms of the dark matter and the dark energy at one hand while the conceptual challenges posed by various black hole space-times and singularity in the theoretical front on the other. In the present work, we have developed a general framework for the Weyl space-time which reduces to the Weyl integral space-time (WIST) for  the Starobinsky modification in $f(R)$ gravity in Einstein-Palatini formalism (EPF). In EPF, the Weyl field turns out to be an undetermined gauge and the space-time turns toWeylian from the Riemannian as soon as the matter is added  to it. \\

The issue of caustics formation is very critical to the location of
occurrence of singularity in certain region of space-time and we
have performed a detailed analysis of this and have applied it, as a
trivial example, to the $f(R)$ gravity cosmological models. 
It is found that the model $R + \beta R^2-2 \Lambda$ clubbed with the Weyl
field does indeed avoid the big-bang singularity which is already
true without the $\beta-$correction. For the Starobinsky modified
gravity along with the suitable initial conditions, it is possible to
have a singularity free cyclic Universe which has unfortunately a 
very short periodicity. It is a pity that the model is not
cosmologically viable, however it offers an interesting example of a
non-singular cyclic model.

\section*{ACKNOWLEDGEMENTS}
RG thanks Centre for Theoretical Physics, Jamia Millia Islamia, New Delhi for hospitality, where this work was initiated. The work of RG is supported by the Grant-in-Aid
for Scientific Research Fund of the JSPS No. 10329. \\


\begin{thebibliography}{99}
\bibitem{Hawking:1973uf}
  S.~W.~Hawking, G.~F.~R.~Ellis,
  Cambridge University Press, Cambridge, 1973.

\bibitem{Wald:1984rg}
  R.~M.~Wald,
  Chicago, USA: Chicago University Press ( 1984) 491p.

\bibitem{Joshi:1987wg}
  P.~S.~Joshi,
  Oxford, UK: Clarendon (1993) 377 p. (International series of monographs on physics, 87).

\bibitem{Ellis:1971pg}
  G.~F.~R.~Ellis,
  Gen.\ Rel.\ Grav.\  {\bf 41 } (2009)  581-660.

\bibitem{Carroll:1997ar}
  S.~M.~Carroll,
  gr-qc/9712019.

\bibitem{Tasi}
  W.~H.~Kinney,
  arXiv:0902.1529 [astro-ph.CO];
  D.~Baumann,
  arXiv:0907.5424 [hep-th].

\bibitem{Fujii:2003pa}
  Y.~Fujii and K.~Maeda,
{\it  Cambridge, USA: Cambridge University Press (2003) 240 p}

\bibitem{Albrecht:1982wi}
  A.~Albrecht, P.~J.~Steinhardt,
  Phys.\ Rev.\ Lett.\  {\bf 48 } (1982)  1220-1223.


\bibitem{clock}
Marzke, R. F. The Theory of Measurement in General Relativity. A.B. senior thesis,
Princeton (1959);\\
Kundt, W., and Hoffmann, B. Recent developments in General Relativity
(Pergamon Press, New York, 1962);\\
Marzke, R. F., and Wheeler, J. A. Gravitation and relativity, (ed. H.Y. Chiu and
W. F. Hoffman). New-York (1964).

\bibitem{Ehlers}
Ehlers, J.; Pirani, F.A.E.; Schild, A. The geometry of free fall and light propagation, in General Relativity, Papers in honour of J.L. Synge, Oxford: Clarendon Press (1972)

\bibitem{Poulis:2011ve}
  F.~P.~Poulis, J.~M.~Salim,
  arXiv:1106.3031 [gr-qc].

\bibitem{Fatibenea:2011aw}
  L.~Fatibenea, M.~Francaviglia,
  arXiv:1106.1961 [gr-qc].

\bibitem{Poisson}E. Poisson, A Relativists Toolkit: The Mathematics of Black Hole Mechanics, Cambridge University Press, 2004.

\bibitem{Anirvan}
A. Dasgupta, H. Nandan and S. Kar, Phys. Rev. {\bf D79} (2009) 124004.

\bibitem{Sardanashvily} G. Sardanashvily and V. Yanchevsky, Acta Physica Polonica {\bf B17} (1986) 1017.

\bibitem{Felder} G. N. Felder, Lev Kofman, and A. Starobinsky, J. High
Energy Phys. {\bf 09} (2002) 026
\bibitem{Goswami} U.D. Goswami, H.Nandan and M. Sami, Phys. Rev. {\bf D82} (2010) 103530


\bibitem{Sayan1} S. Kar and S. Sengupta, Pramana {\bf 69} (2007) 49.

\bibitem{Dadhich} N. Dadhich, Pramana {\bf 69} (2007) 23; N Dadhich, gr-qc/0511123.

\bibitem{Anirvan1}
A. Dasgupta, H. Nandan and S. Kar, Annals Phys. {\bf 323} (2008) 1621;
A. Dasgupta, H. Nandan and S. Kar, Int. J. Geom. Meth. Mod. Phys. {\bf 6} (2009)
645, Erratum-ibid. {\bf 07} (2010) 517; S. Ghosh, A. Dasgupta and S. Kar, Phys. Rev.
{\bf D83} (2011) 084001.


\bibitem{Starobinsky:1980te}
  A.~A.~Starobinsky,
  Phys.\ Lett.\  {\bf B91 } (1980)  99-102.

\bibitem{Carroll:2003wy}
  S.~M.~Carroll, V.~Duvvuri, M.~Trodden and M.~S.~Turner,
  Phys.\ Rev. {\bf D70} (2004) 043528
  [arXiv:astro-ph/0306438].

\bibitem{Poberii:1994rz}
  E.~A.~Poberii,
  Gen.\ Rel.\ Grav.\  {\bf 26 } (1994)  1011-1054.

\bibitem{Schouten}
J. A. Schouten, Ricci Calculus (2nd ed.; Berlin, 1954)

\bibitem{Brown:1999fx}
  H.~R.~Brown and O.~Pooley,
  arXiv:gr-qc/9908048.

\bibitem{Novello:1992tb}
  M.~Novello, L.~A.~R.~Oliveira, J.~M.~Salim, E.~Elbaz,
  Int.\ J.\ Mod.\ Phys.\  {\bf D1 } (1992)  641-677.

\bibitem{arXiv:0912.0432}
  T.~Y.~Moon, J.~Lee and P.~Oh,
  Mod.\ Phys.\ Lett.\ A\ {\bf 25} (2010) 3129
  [arXiv:0912.0432 [gr-qc]].
  
\bibitem{Novello:2008ra}
  M.~Novello and S.~E.~P.~Bergliaffa,
  Phys.\ Rept.\  {\bf 463} (2008) 127
  [arXiv:0802.1634 [astro-ph]].

\bibitem{Weyl}
H. Weyl, Space Time Matter, Dover

\bibitem{Pauli}
W. Pauli, Theory of Relativity (Pergamon, London, 1958)

\bibitem{Salim:1996ei}
  J.~M.~Salim and S.~L.~Sautu,
  Class.\ Quant.\ Grav.\  {\bf 13} (1996) 353.

\bibitem{Cotsakis:1997cj}
  S.~Cotsakis, J.~Miritzis and L.~Querella,
  J.\ Math.\ Phys.\  {\bf 40} (1999) 3063
  [arXiv:gr-qc/9712025].

\bibitem{Sotiriou:2005cd}
  T.~P.~Sotiriou,
  Class.\ Quant.\ Grav.\  {\bf 23} (2006) 1253
  [arXiv:gr-qc/0512017].
\end{thebibliography}
\end{document}